\documentclass[12pt]{article}
\begin{document} 
~~~~~~~~~~~~~~~~~~~~~~~~~~~~~~~~~~~~~~~~~~~Preprint-INR-TH-044

\begin{center}
{\huge \bf Higher-derivative relativistic quantum gravity} \\[10mm] 
 S.A. Larin \\ [3mm]
 Institute for Nuclear Research of the
 Russian Academy of Sciences,   \\
 60th October Anniversary Prospect 7a,
 Moscow 117312, Russia
\end{center}

\vspace{30mm}

\begin{abstract} 
Relativistic quantum gravity with the action including terms quadratic
in the curvature tensor is analyzed. We derive new expressions for the
corresponding Lagrangian and the graviton propagator within dimensional
regularization. We argue that the considered model is  a good candidate
for the fundamental quantum theory of gravitation. 
\end{abstract}

\newpage

Creation of the fundamental quantum theory of gravitation remains
one of the most important tasks, if not he most important task, of mofern
theoretical physics. It is known that the problem arises because
of non-renormalizability \cite{h} of General Relativity.

Quite long ago Stelle has proved renormalizability of the 
Lorentz invariant gravitational
actions which include besides the Einstein-Hilbert term also terms with
fourth derivatives of the metric \cite{st}. His proof used the spesific
gauge where the structure of ultraviolet divergences is particularly
simple. For more general gauges he made the hypothesis of the cohomological
structure of divergences. Recently this hypothesis was proved for the 
general class of background gauges \cite{sib}. Thus we consider the fact
of  renormalizability of  garvity
with fourth derivatives of the metric, which we will call also 
quadratic quantum gravity, as well established. 

But Stelle also made the statement \cite{st,st2} that quantum gravity with
fourth derivatves is unphysical because it violates either unitarity 
or causality and the model can serve only as an effective theory in some
domain of energies. Since then this model is considered as having severe
problems with physical interpretation.

In the present paper we derive new expressions for the Lagrangian
and for the graviton propagator of quadratic quantum gravity within
dimensional regularization. We argue also that fourth derivative gravity
is a good candidate for the fundamental quantun theory of gravitation. 

Let us consider the invariant under the gauge 
transformations action  with all possible terms quadratic in the
curvature tensor
\begin{equation}
\label{action}
S_{sym}=\int d^{D}x \mu^{-2\epsilon} \sqrt{-g} \left(-M_{Pl}^2 R 
+\alpha R_{\mu\nu}R^{\mu\nu}
+\beta R^2+\delta R_{\mu\nu\rho\sigma}R^{\mu\nu\rho\sigma}\right),
\end{equation}
where the first term is the Einstein-Hilbert action.
The $\Lambda$-term is omitted
in the action since we will work within peturbation theory.
Here $M_{Pl}^2=1/(16\pi G)$ is the squared Planck mass,
$R_{\mu\nu\rho\sigma}$ is the Riemann tensor, $R_{\mu\nu}$ is the Ricci tensor,
$R$ is the Ricci scalar, $\alpha$, $\beta$ and $\delta$ are the dimensionless
coupling constants of the Lagrangian, $D=4-2\epsilon$ is the space time dimension within 
dimensional regularization, $\mu$ is the parameter of dimensional regularization.

We would like to stress that dimensional regularization is presently the only
practically available regularization to regularize ultraviolet divergencies
preserving gauge invariance in quantum gravity.

Usually the last term in the action (\ref{action}) is missed in the literature
\cite{st,sib,str}  because of  
the Gauss-Bonnet topological identity
\begin{equation}
\int d^4x  \sqrt{-g}\left( 
R_{\mu\nu\rho\sigma}R^{\mu\nu\rho\sigma}-4R_{\mu\nu}R^{\mu\nu}+R^2\right)=0,
\end{equation}
which is valid for space-times topologically equivalent to flat
 space only in four dimensions.
Within dimensional regularization the term quadratic in the 
Riemann tensor should be added to the ation.

We work in the linearized theory around the flat space metric
\begin{equation}
g_{\mu\nu}=\eta_{\mu\nu}+h_{\mu\nu},
\end{equation}
where we choose the convention
$\eta_{\mu\nu}=diag(+1,-1,-1,-1)$ in four dimensions. 
In $D$ dimensions  $\eta_{\mu\nu}\eta^{\mu\nu}=D$.
Further it is understood that indices are raised and lowered with
the Minkowski metric  $\eta_{\mu\nu}$.

Gauge transformation are generated by diffeomorphisms
$x^{\mu} \rightarrow x^{\mu}+\zeta^{\mu}(x)$ and have the form
\begin{equation}
h_{\mu\nu}\rightarrow h_{\mu\nu}+\partial_{\mu}\zeta_{\nu}+
\partial_{\nu}\zeta_{\mu}
+\left(h_{\lambda\mu}\partial_{\nu}+h_{\lambda\nu}\partial_{\mu}+
(\partial_{\lambda} h_{\mu\nu})\right)\zeta^{\lambda},
\end{equation}
here $\zeta_{\mu}(x)$ are arbitrary functions.

According to Faddeev-Popov quantization one should add to the action the 
gauge fixing term
which we choose in the form
\begin{equation}
S_{gf}=-\frac{1}{2\xi}\int d^D x F_{\mu}
\partial_{\nu}\partial^{\nu} F^{\mu},
\end{equation}
where $F^{\mu}=\partial_{\nu}h^{\nu\mu}$, $\xi$ is the gauge parameter.

One should also add the ghost term 
\begin{equation}
S_{ghost}=\int d^Dx d^D y \overline{C}_{\mu}(x)
\frac{\delta F^{\mu}(x)}{\delta \zeta_{\nu}(y)}C_{\nu}(y)=
\end{equation}
\[ 
\int d^D x \partial^{\nu}\overline{C^{\mu}}\left[\partial_{\nu}C_{\mu}
+\partial_{\mu}C_{\nu}
+h_{\lambda\mu}\partial_{\nu}C^{\lambda}
+h_{\lambda\nu}\partial_{\mu}C^{\lambda}+
(\partial_{\lambda}h_{\mu\nu})C^{\lambda}\right], 
\]
where $\overline{C}$ and $C$ are ghost fields.
Thus one gets the following generating functional for Green functions 
of gravitons
\begin{equation}
Z(J)=N \int d h_{\mu\nu} dC_{\lambda} d\overline{C_{\rho}}
 \exp{\left[i\left(S_{sym} +S_{gf}+S_{ghost}+
d^Dx \mu^{-2\epsilon} J_{\mu\nu} h^{\mu\nu}\right)\right]},
\end{equation}
where as usual in the functional integral, $N$ is the normalization factor 
and $J_{\mu\nu}$
is the source of the gravitational field.

We work within perturbation theory, so we make the shift of the fields
\begin{equation}
 h_{\mu\nu} \rightarrow M_{Pl} \mu^{-\epsilon} h_{\mu\nu}.
\end{equation}
Thus perturbative expansion goes in the invers powers of the Plank mass 
or, in other words,
 in the powers of the Newton coupling constant $G$.

To derive the graviton propagator we make the Fourier transfom 
to the momentum space 
and write the quadratic in $ h_{\mu\nu}$ form
\[
Q_{\mu\nu\rho\sigma}=
\frac{1}{4}\int d^D k~ h^{\mu\nu}(-k)\left[\left(k^2+M_{Pl}^{-2}k^4(\alpha 
+4\delta)\right)P^{(2)}_{\mu\nu\rho\sigma} \right.
\]
\begin{equation}
\label{forma}
+k^2\left(-2+4M_{Pl}^{-2}k^2(\alpha+3\beta+\delta)\right)
P^{(0-s)}_{\mu\nu\rho\sigma}
\end{equation}
\[ \left. 
+\frac{1}{\xi}M_{Pl}^{-2}k^4\left(P^{(1)}_{\mu\nu\rho\sigma}
+2P^{(0-w)}_{\mu\nu\rho\sigma}\right) \right] h^{\rho\sigma}(k),
\]
where $P^{(i)}_{\mu\nu\rho\sigma}$ are projectors 
to the spin-2, spin-1 and spin-0 
components of the field 
 correspondingly:
\begin{equation}
P^{(2)}_{\mu\nu\rho\sigma}=\frac{1}{2}\left(\Theta_{\mu\rho}\Theta_{\nu\sigma}
+\Theta_{\mu\sigma}\Theta_{\nu\rho}\right)
-\frac{1}{3}\Theta_{\mu\nu}\Theta_{\rho\sigma},
\end{equation}
\begin{equation}
P^{(1)}_{\mu\nu\rho\sigma}=\frac{1}{2}\left(\Theta_{\mu\rho}\omega_{\nu\sigma}
+\Theta_{\mu\sigma}\omega_{\nu\rho}
+\Theta_{\nu\rho}\omega_{\mu\sigma}+\Theta_{\nu\sigma}\omega_{\mu\rho}\right),
\end{equation}
\begin{equation}
P^{(0-s)}_{\mu\nu\rho\sigma}=\frac{1}{3}\Theta_{\mu\nu}\Theta_{\rho\sigma},
\end{equation}
\begin{equation}
P^{(0-w)}_{\mu\nu\rho\sigma}=\omega_{\mu\nu}\omega_{\rho\sigma}.
\end{equation}
Here $\Theta_{\mu\nu}=\eta_{\mu\nu}-k_{\mu}k_{\nu}/k^2$ and 
$\omega_{\mu\nu}=k_{\mu}k_{\nu}/k^2$
are the transvers and longitudinal projectors for vector quantities.

The sum of the projectors is the unity
\begin{equation}
P^{(2)}_{\mu\nu\rho\sigma}+P^{(1)}_{\mu\nu\rho\sigma}
+P^{(0-s)}_{\mu\nu\rho\sigma}+P^{(0-w)}_{\mu\nu\rho\sigma}=
\frac{1}{2}(\eta_{\mu\rho}\eta_{\nu\sigma}
+\eta_{\mu\sigma}\eta_{\nu\rho}).
\end{equation}

Note that the expression (\ref{forma}) essentially differs from the analogous
one in ref. \cite{str} by the absence of $\epsilon$-dependent contributions.

To obtain the graviton propagator $D_{\mu\nu\rho\sigma}$
one inverts the matrix in the square brackets of (\ref{forma}):
\begin{equation}
[Q]_{\mu\nu\kappa\lambda}D^{\kappa\lambda\rho\sigma}
=\frac{1}{2}(\delta_{\mu}^{\rho}\delta_{\nu}^{\sigma}
+\delta_{\mu}^{\sigma}\delta_{\nu}^{\rho}).
\end{equation}
Thus we get for the propagator
\begin{equation}
D_{\mu\nu\rho\sigma}=\frac{1}{i(2\pi)^D}
\left[
\frac{4}{k^2}\left(\frac{1}{1+M_{Pl}^{-2}k^2(\alpha +4\delta)}\right)
P^{(2)}_{\mu\nu\rho\sigma}
\right.
\end{equation}
\[
-\frac{2}{k^2}\left(\frac{1+2\epsilon\frac{1-M_{Pl}^{-2}k^2(\alpha+4\beta 
 )}{1+M_{Pl}^{-2}k^2(\alpha+4\delta)}}
 {1-\epsilon-M_{Pl}^{-2}k^2\left((2\alpha+6\beta+2\delta)
-\epsilon (\alpha+4\beta)\right)}\right)
P^{(0-s)}_{\mu\nu\rho\sigma}
\]
\[ \left.
+4\xi \frac{1}{ M_{Pl}^{-2}k^4} \left(P^{(1)}_{\mu\nu\rho\sigma}+
\frac{1}{2}P^{(0-w)}_{\mu\nu\rho\sigma}\right)\right].
\]
Let us perform partial fractioning. Then the graviton propagatr
takes the form
\begin{equation}
D_{\mu\nu\rho\sigma}=\frac{1}{i(2\pi)^D}
\left[
4P^{(2)}_{\mu\nu\rho\sigma}\left(\frac{1}{k^2}
-\frac{1}{k^2-M_{Pl}^2/(-\alpha-4\delta)}\right)
\right. 
\end{equation}
\[ 
-2\frac{P^{(0-s)}_{\mu\nu\rho\sigma}}{1-\epsilon}\left(
1+2\epsilon\frac{1-M_{Pl}^{-2}k^2(\alpha+4\beta)}
{1+M_{Pl}^{-2}k^2(\alpha+4\delta)}\right)
\]
\[
\left(\frac{1}{k^2}-\frac{1}{k^2-M_{Pl}^2(1-\epsilon)/
(2\alpha +6\beta+2\delta-\epsilon (\alpha+4\beta))}\right)
\]
\[ \left.
+\frac{4\xi}{M_{Pl}^{-2}k^4}\left(P^{(1)}_{\mu\nu\rho\sigma}+
\frac{1}{2} P^{(0-w)}_{\mu\nu\rho\sigma}\right)\right].
\]

It is interesting to note that the position of one of the poles in the term
with $P^{(0-s)}_{\mu\nu\rho\sigma}$ depends on the regularization
parameter $\epsilon$. The residues of both poles in this term also depend
on  $\epsilon$. Thus it is clear that poles and residues of the tree
level propagator do not have direct physical meaning.

In the limit of four dimensions we get for the  graviton propagator
\begin{equation}
\label{four}
D_{\mu\nu\rho\sigma}=\frac{4}{i(2\pi)^D}
\left[
\frac{P^{(2)}_{\mu\nu\rho\sigma}
-\frac{1}{2}P^{(0-s)}_{\mu\nu\rho\sigma}}{k^2}
-\frac{P^{(2)}_{\mu\nu\rho\sigma}}{k^2-M_{Pl}^2/(-\alpha-4\delta)}
\right.
\end{equation}
\[ \left.
+\left(\frac{1}{2}\right)\frac{P^{(0-s)}_{\mu\nu\rho\sigma}}{k^2-M_{Pl}^2/
(2\alpha +6\beta+2\delta)}
+\frac{\xi}{M_{Pl}^{-2}k^4}\left(P^{(1)}_{\mu\nu\rho\sigma}+
\frac{1}{2} P^{(0-w)}_{\mu\nu\rho\sigma}\right)\right],
\]

Within classical four-derivative gravity   
for a point particle 
with the energy-momentum tensor
$T_{\mu\nu}=\delta^0_{\mu}\delta^0_{\nu}M\delta^3(x)$ 
the gravitational field has the form \cite{st2} 
\begin{equation}
\label{potential}
V(r)=\frac{M}{2 \pi M_{Pl}^2}\left(-\frac{1}{4 r}+\frac{e^{-m_{2} r}}{3 r}
-\frac{e^{-m_0 r}}{12 r}\right),
\end{equation}
where in our notations $m_2^2= M_{Pl}^2/(-\alpha-4\delta)$
and  $m_0^2=M_{Pl}^2/(2\alpha+6\beta+2\delta)$ are the squared masses
of the massive spin-2 and spin-0 gravitons correspondingly. Naturally
the values of $\alpha, \beta$ and $\delta$ should be chosen to ensure
positivity of the masses. 
As it was mentioned 
in \cite{st,st2} these masses can be chosen large enough 
to have agreements with present experiments.

The propagator (\ref{four}) reproduces 
this expression after the calculation of
the corresponding tree level Feynman diagram describing interaction of two
point-like particles. The propagator presented in \cite{st} contains some
errors and dos not reproduce the expression (\ref{potential}). 

The second term in the graviton propagator (\ref{four}) has the 
unusial minus sign
and that is why it is interpreted as the massive spin-2 ghost.
To preserve renormalizability of the quantum theory one should shift
all poles in propagators in Feynman integrals in the same manner
$k^2 \rightarrow k^2 +i0$, hence the ghost state should be considered
as the state with the negative metric \cite{st}.
This was the reason to make the statement \cite{st,st2} about violation
either unitarity or causality in the model.

But this massive spin-2 state is unstable since it can decay into
massless gravitons. Thus it does not appear as the asymptotic state
of the $S$-matix. Correspondingly only particles with the positive metric
participate in the scattering processes as external particles
and unitarity is preserved
in the theory.

Unitarity of the $S$-matrix in the presence
of negative metric states was considered in particular
 in \cite{lee,cut} where the
question of causality was also studied.

It should be also mentioned that the tree level 
propagator (\ref{four}) will be 
essentially modified after the summation of one-loop corrections.
Because of the mignus sign in the second term of (\ref{four})
the one-loop correction due to the diagram with the massless graviton
in the loop will shift the pole of the ghost from the real value
 $k^2= M_{Pl}^2/(-\alpha-4\delta)$ to the complex value
$k^2= M_{Pl}^2/(-\alpha-4\delta)-i\Gamma$, where $\Gamma$
is the decay width of the massive spin-2 graviton into the pare
of massless gravitons.
The complex pole is located on the
unphysical Riemann sheet. This is analogous to the known
virtual level in the neutron-proton system 
with antiparallel spins \cite{ll}.

We have considered only purely gravitational action. But us it was
shown in \cite{st} one can include the matter fields in the theory
straightforwardly.

We conclude that the considered quadratic quantum gravity is a good
candidate for fundamental quantum theory of gravittion.

The author is grateful to the collaborators of the Theory division of INR
D. Gorbunov, M. Ivanov, D. Legkov,  M. Libanov,  E. Nugaev,  V. Rubakov 
S. Sibiryakov and A. Smirnov
for helpful discussions.

\end{document}